\documentclass [10pt,compsocconf]{IEEEtran}
\usepackage{algorithm,algorithmic,amsbsy,amsmath,amssymb,epsfig,bbm,mathrsfs,multirow,amsthm}
\usepackage{subfigure,float}

\usepackage[top=0.75in, bottom=0.67in, left=0.67in, right=0.67in]{geometry}

\DeclareMathOperator*{\argmax}{argmax}

\newcommand\blfootnote[1]{%
	\begingroup
	\renewcommand\thefootnote{}\footnote{#1}%
	\addtocounter{footnote}{-1}%
	\endgroup
}

\begin{document}
	\title{Optimal Cross Slice Orchestration for 5G Mobile Services}
	\author{Dinh Thai Hoang$^1$, Dusit Niyato$^1$, Ping Wang$^1$, Antonio De Domenico$^2$ and Emilio Calvanese Strinati$^2$ \\
		$^1$ School of Computer Science and Engineering, Nanyang Technological University, Singapore\\
		$^2$  CEA, LETI, MINATEC, F-38054 Grenoble, France	\vspace{-5mm}	}
	
	\maketitle	
	\blfootnote{Part of this work has been performed within the 5G-MoNArch project, part of the Phase II of the 5th Generation Public Private Partnership (5G PPP) and partially funded by the European Commission within the Horizon 2020 Framework Programme.}
	\begin{abstract}
		5G mobile networks encompass the capabilities of hosting a variety of services such as mobile social networks, multimedia delivery, healthcare, transportation, and public safety. Therefore, the major challenge in designing the 5G networks is how to support different types of users and applications with different quality-of-service requirements under a single physical network infrastructure. Recently, network slicing has been introduced as a promising solution to address this challenge. Network slicing allows programmable network instances which match the service requirements by using network virtualization technologies. However, how to efficiently allocate resources across network slices has not been well studied in the literature. Therefore, in this paper, we first introduce a model for orchestrating network slices based on the service requirements and available resources. Then, we propose a Markov decision process framework to formulate and determine the optimal policy that manages cross-slice admission control and resource allocation for the 5G networks. Through simulation results, we show that the proposed framework and solution are efficient not only in providing slice-as-a-service based on the service requirements, but also in maximizing the provider's revenue. 
	\end{abstract}

	{\it Keywords-} 5G networks, network slicing, Markov decision processes, admission control. 
	
	\section{Introduction}
	The fifth generation (5G) mobile network is currently attracting tremendous research interest from both industry and academia due to its significant benefits and huge market potential. Compared to the current 4G network, the 5G network is expected to achieve 1,000 times higher system throughput, 10 times higher spectral efficiency and data rates (i.e., the peak data rate of 10 Gb/s and the user experienced rate of 1Gb/s), 5 times reduction in end-to-end latency, and 100 times higher connectivity density~\cite{Wang2014Cellular}. In addition, different from 4G networks where all mobile users are served by a communication network, 5G networks need to tailor on diverse mobile services with different demands and requirements. Thus, network slicing technique has been emerging as an enabling solution that allows mobile 5G network providers to achieve such goals.
	
\begin{figure}[!]
		\begin{center}
			\epsfxsize=3.5 in \epsffile{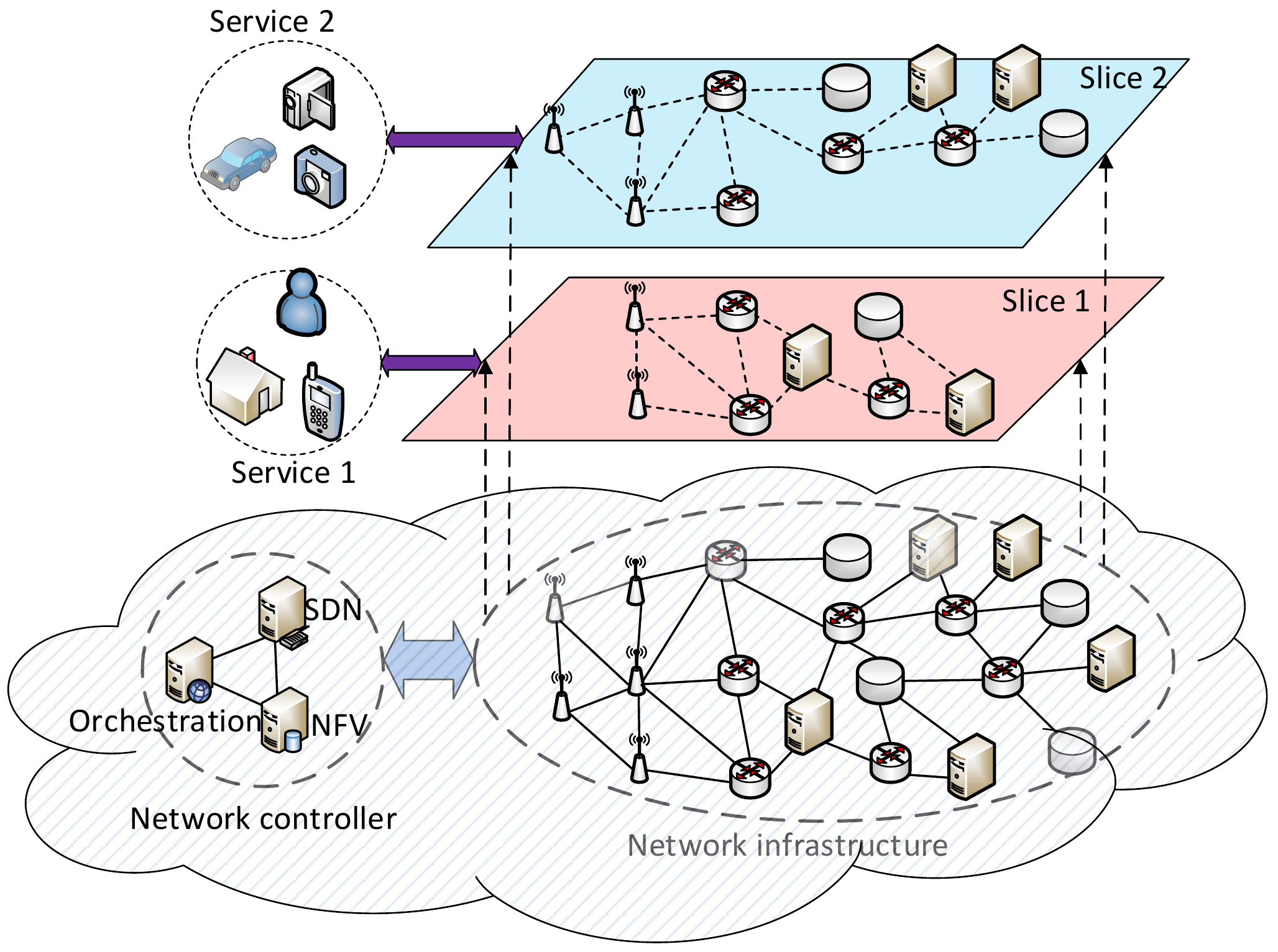} 
			\caption{Architecture of 5G networks with slicing.}
			\label{fig:Architecture}
		\end{center}
		\vspace{-0.5cm}
	\end{figure}

	Specifically, network slicing is a new network virtualization technique that splits a single physical infrastructure into multiple virtual networks, i.e., slices (as illustrated in Fig.~\ref{fig:Architecture}), with functionalities designed to serving specific demands and requirements~\cite{Nikaein2015Network}. The core idea of the network slicing is using software defined networking (SDN) and network functions virtualization (NFV) technologies for virtualizing the physical infrastructure and controlling network operations. In particular, SDN provides a separation between the network control and data planes, improving flexibility of network function management and efficiency of data transfer. NFV allows various network functions to be virtualized, i.e., in virtual machines. As a result, the functions can be moved to different locations and the corresponding virtual machines can be migrated to run on commoditized hardware dynamically depending on the demand and requirements. As such, SDN will play a significant role in the control of the NFV infrastructure resources (both physical and virtual) by enabling automatic network configuration and policy control.
	
	The key benefit of network slicing is to enable providers to offer network services on an as-a-service basis which enhances operational efficiency while reducing time-to-market for new services~\cite{Zhou2016Network}. However, orchestrate the slice requests and manage the network resources are open challenges. Thus, optimization techniques can be adopted to find decisions for the provider given the service demands and available resources. In this paper, we first introduce a system model which groups network slices based on their demands. Then, we formulate the cross slice admission control problem as a Markov decision process (MDP) and adopt the value iteration algorithm to find the optimal policy for the provider. Through simulation results, we demonstrate that the proposed model and solution can achieve the best performance in term of average reward. In addition, the simulation results also reveal impacts of parameters, e.g., arrival and departure probabilities of requests, to the system performance. This information is especially important to the provider in controlling quality-of-service as well as in maximizing its profits. 
	

	\section{Related Work}
	\label{sec:Rel-Work}
	
	There are some research works related to designing, controlling, and orchestrating network slicing in 5G networks. In~\cite{Yoo2016Network}, the author discussed design issues of network slicing in 3GPP networks, and introduced a new network slicing architecture to address three design problems including standardization, network slice selection, and slice-independent functions. The core idea of the proposed architecture is based on the top-down design concept and the proposal of NextGen radio resource control. However, the proposed architecture has a high signaling overhead due to many exchanged messages. 
	
	The authors in~\cite{Choyi2016Network} presented a framework for providing customized network slices in 5G networks based on Quality-of-Service identifier (QCI) and security requirements. In this framework, a network slice will be allocated to the requested user based on a service description document which contains details of the services and their corresponding QCI, e.g., latency, throughput, and security level. Moreover, this framework allows the user to negotiate with the service provider to choose the best service meeting the user's demands. Service-based slice selection was also studied in~\cite{Sama2016Service}. However, in~\cite{Sama2016Service} the network slice selection is based on the requirements of service groups rather than on the requirements of each individual user. In some cases in practice, the requirements from different users are overlapped, and thus the requirements can be naturally grouped into a service group (as illustrated in Fig.~\ref{fig:Architecture}). Under the proposed solution, network efficiency and revenue for the service provider can be greatly improved. 
	
	Spectrum sharing among slices with the same air interface is a challenge because spectrum must be allocated not only to meet the users' demands, but also to achieve the best performance without interference between slices. Although frequency-division multiple access technique provides the best isolation, it may result in resource under-utilization due to the loss of statistical multiplexing gain~\cite{GENITechnical2006}. Therefore, the authors in~\cite{Soliman2016QoS} examined an approach using space-division multiple-access technique to share spectrum resource among slices according to the frequency and space dimensions, while taking into account the performance difference between frequency and spatial multiplexing. This solution can improve network throughput for the provider, but it requires precoding processes which may cause delay for providing services. 
	
	The closest work with our paper is~\cite{Jiang2016Network} where the resource allocation problem for 5G networks using network slicing was studied. However, unlike~\cite{Jiang2016Network} where the authors just focused on spectrum resource allocation problem to meet the users' Quality-of-Services (QoSs), in this paper, we jointly consider the computing, storage, and spectrum resources in allocating slices to services. Moreover, different from~\cite{Jiang2016Network} where the resources of slices are fixed and predetermined, in this paper, we proposed a dynamic cross-slice admission control scheme to allow providers to supply flexible services according to the service demands. To the best of our knowledge, this is the first work which proposes using the MDP framework to find the optimal cross-slice orchestration policy in 5G networks.

	\section{System Model}
	\label{sec:System Model}
	
\begin{figure}[!h]
		\begin{center}
			\epsfxsize=3.5 in \epsffile{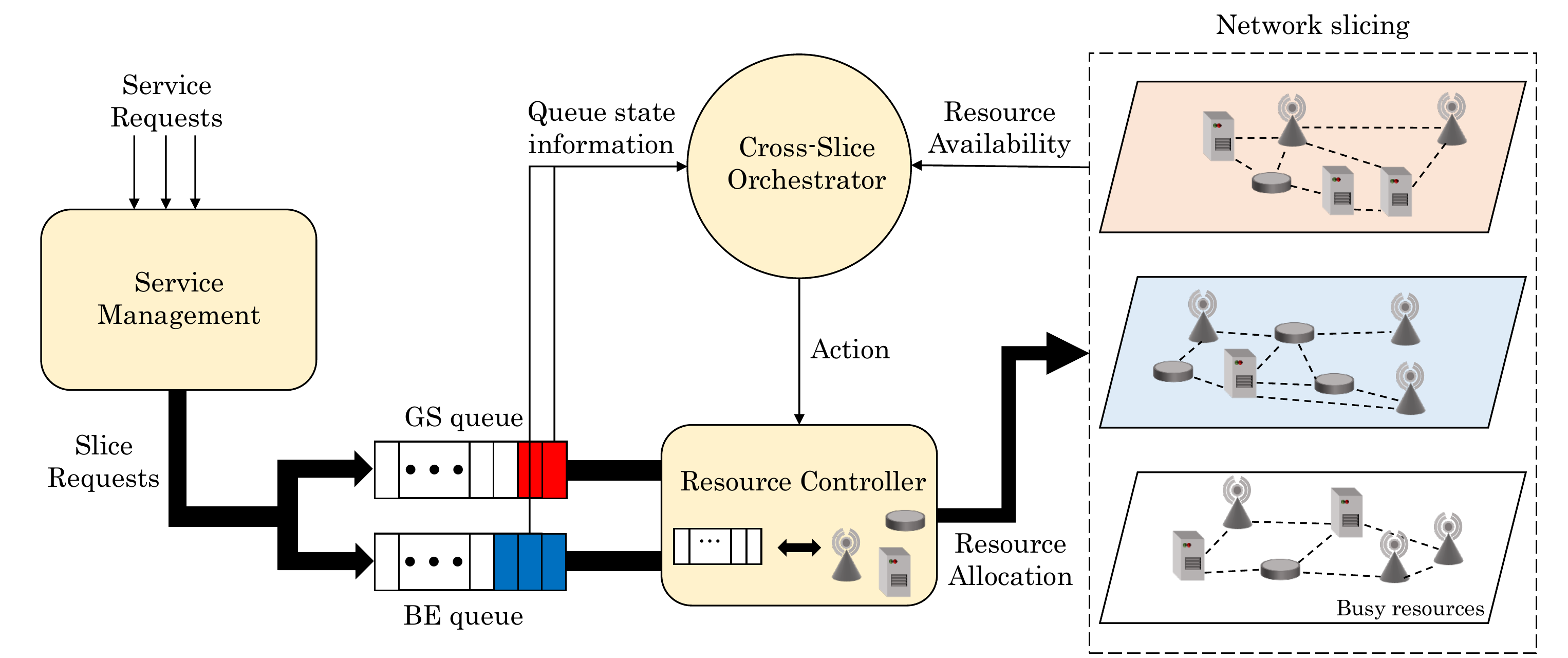} 
			\caption{Proposed System Model.}
			\label{fig:System_Model}
		\end{center}
		\vspace{-0.2cm}
	\end{figure}

	Fig.~\ref{fig:System_Model} describes the system model of a 5G network with network slicing. In this model, when network service requests arrive at the Service Management block, they are mapped to slices with specific service requirements, then analyzed and classified into two types of requests, i.e., guaranteed QoS (GS) and best effort (BE) slices. For example, GS requests are related to virtual reality services, while BE requests are associated with more classical mobile broadband. The slice requests are then stored and transferred to the corresponding queues. At the end of each time slot, the requests that have not been sent to the queues will be removed from the buffer. Then, based on the requests in the two queues together with the current state of the available network resources, the Cross-Slice Resource Orchestrator makes a decision to choose the slice to be admitted in the system. This decision is sent to the Resource Controller that instantiates slices by allocating the required physical resources.
	\vspace{-0.3cm}

\begin{figure}[h]
		\begin{center}
			\epsfxsize=3.5 in \epsffile{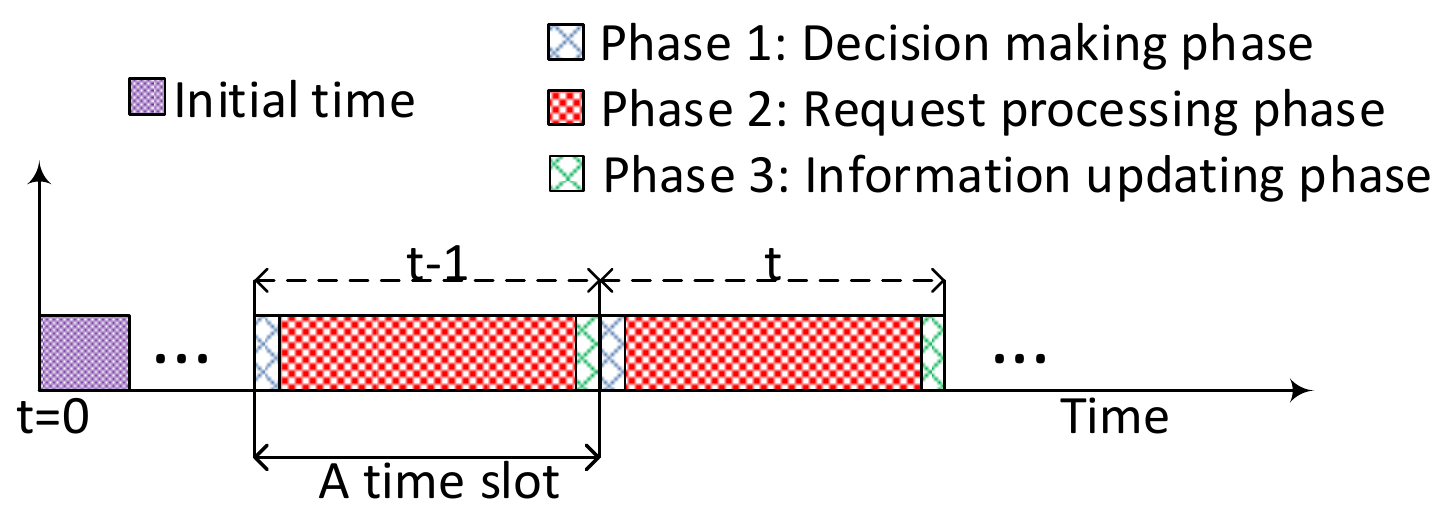} 
			\caption{Proposed Time frame structure.}
			\label{fig:TimeFrameStructure}
		\end{center}
	\end{figure}

	Based on the proposed system model, each time slot can be divided into three phases as illustrated in Fig.~\ref{fig:TimeFrameStructure}. At the beginning of each time slot, i.e., the decision making phase, based on the current states of the two queues, and the current available resources of the system, the Cross-Slice Orchestrator will choose the slice requests to admit. Then, in the second phase, i.e., the request processing phase, the Resource Controller will allocate network resources to the selected requests. Finally, in the last phase, i.e., the information updating phase, the Cross-Slice Orchestrator updates the information related to the active queues and available network resources.
	
	At each time slot, we assume that there are $n_g$ and $n_b$ slice requests for GS and BE services arriving at the system, respectively. We denote $n_g \in \mathcal{N}_g=\{0,1,\ldots,N_g\}$ and $n_b \in \mathcal{N}_b=\{0,1,\ldots,N_b\}$ where $N_g$ and $N_b$ are the maximum numbers of arriving requests for the GS and BE services in one time slot, respectively. In addition, we denote $p_n^g$ and $p_n^b$ as the probabilities that there are $n_g$ and $n_b$ requests arriving at the system in one time slot. Then, we have 
	\begin{equation}
	\sum_{n=0}^{N_g} p_n^g = 1 \phantom{10} \text{and} \phantom{10} \sum_{n=0}^{N_b} p_n^b = 1.
	\end{equation}
	Similarly, we denote $p_l^g \in (0,1]$ and $p_l^b \in (0,1]$ as the probabilities which a running slice ends in the current time slot. Here, we note that after a slice life cycle is completed, its request will be removed from the system, and at the same time the corresponding resources will be released.

	\section{Optimization Formulation and Solution}
	\label{sec:Opt_For}
	
	In this section, we first formulate the cross-slice admission control and resource allocation optimization problem as an MDP, and then present the value iteration method to find the corresponding optimal policy. 
	
	\subsection{State Space}
	
	The state space of the system, denoted by $\mathcal{S}$, includes the states of the two queues and the state of the available network resources, which are observed at the Cross-Slice Orchestrator, and is defined as follows:
	\begin{equation}
	\mathcal{S} \triangleq \mathcal{S}_g \times \mathcal{S}_b \times \mathcal{S}_p	,
	\end{equation}
	where $\mathcal{S}_g$, $\mathcal{S}_b$, and $\mathcal{S}_p$ are the state spaces of GS queue, BE queue, and network resources, respectively. If we denote $s_g$, $s_b$, and $s_p$ as the state of GS queue, BE queue, and network resources, respectively, the composite state of the system can be represented by $s=(s_g,s_b,s_p)$. Note that $s_g \in \mathcal{S}_g = \{0,1,\ldots, Q_g\}$ and $s_b \in \mathcal{S}_b = \{0,1,\ldots, Q_b\}$, where $Q_g$ and $Q_b$ are the maximum queue lengths of the GS and BE queues, respectively. 
	
	In the system under consideration, the network resources include radio, computing, and storage resources. Thus, if we denote $r$, $c$, and $\delta$ as the states of the available radio, computing, and storage resources, respectively, we can define $s_p = (r,c,\delta)$. Let denote $R$, $C$, and $\Delta$ as the maximum number of available radio resource units, computing resource units, and storage resource units. We have $r \in \{0,1,\ldots,R\}$, $c \in \{0,1,\ldots,C\}$, and $\delta \in \{0,1,\ldots,\Delta\}$. 
	
	\subsection{Action Space}
	
	In our considered system, at each time slot, the Cross-Slice Orchestrator has to decide how many GS and BE slice requests waiting in the queues will be admitted. Hence, if we denote $a_g$ and $a_b$ as the number of chosen GS and BE requests, action $a$ and the action space $\mathcal{A}$ of the Cross-Slice Orchestrator are defined as follows:
	\begin{equation}
	{\mathcal{A}} \triangleq \{a = (a_g,a_b) \} .
	\end{equation}
	
	The system state may change over the time slots, and thus the action at each time slot must be selected based on the current system state under the following constraints:
	\begin{equation}
	\label{eq:action_constrain1}
	a_g(t) \leq s_g(t) \phantom{10} \text{and} \phantom{10} a_b(t) \leq s_b(t)	,
	\end{equation}
	and
	\begin{equation}
	\label{eq:action_constrain2}
	a_g(t) d_r^g + a_b(t) d_r^b \leq r(t) ,
	\end{equation}
	\begin{equation}
	\label{eq:action_constrain3}
	a_g(t) d_c^g + a_b(t) d_c^b \leq c(t) ,
	\end{equation}
	\begin{equation}
	\label{eq:action_constrain4}
	a_g(t) d_{\delta}^g + a_b(t) d_{\delta}^b \leq {\delta}(t) ,
	\end{equation}
	where $d_r^g$, $d_c^g$, and $d_{\delta}^g$ are the number of units of radio, computing, and storage resources, respectively, required by a GS slice request. Similarly, $d_r^b$, $d_c^b$, and $d_{\delta}^b$ are the number of units of radio, computing, and storage resources, respectively, required by a BE slice request. Eq.~(\ref{eq:action_constrain1}) means that the number of admitted slices cannot exceed the number of requests waiting in the queues. The conditions in~(\ref{eq:action_constrain2}),~(\ref{eq:action_constrain3}), and~(\ref{eq:action_constrain4}) ensure that the resources required by the admitted slices do not exceed the current available resources of the system.
	
	\subsection{Transaction Probability Matrix}
	
	We first express the transition probability matrix given action $a \in {\mathcal{A}}$ as follows:
	\begin{equation}
	{\mathbf{P}}(a) \! =	\!  \left[\!	\begin{array}{c@{\hspace{0.3em}}c@{\hspace{0.3em}}c@{\hspace{0.3em}}c}
	{\mathbf{B}}_{0,0}(a)	&	{\mathbf{B}}_{0,1}(a) &	\ldots & {\mathbf{B}}_{0,Q_b}(a) 	\\
	{\mathbf{B}}_{1,0}(a)	&	{\mathbf{B}}_{1,1}(a)	&	\ldots  & {\mathbf{B}}_{1,Q_b}(a)	\\
	\vdots &	\vdots	&	\ddots	&	\vdots		\\
	{\mathbf{B}}_{Q_b,0}(a) & {\mathbf{B}}_{Q_b,1}(a)	&	\ldots	&	{\mathbf{B}}_{Q_b,Q_b}(a)
	\end{array}\!	\right]
	\begin{array}{l}	\! \leftarrow	b=0	\\	\! \leftarrow	b=1	\\	\vdots	\\	\! \leftarrow	b = Q_b \end{array}		
	\label{eq:pmatrix_B}
	\end{equation}
	where each row of matrix ${\mathbf{P}}(a)$ corresponds to the number of requests in the BE queue. The matrix ${\mathbf{B}}_{b,b'}(a)$ represents the queue state transition probability from state $b$ in the current time slot to state $b'$ in the next time slot given action $a$. This probability depends on the BE request arrival and the selected action. For example, if $N_b=1$, the current state $s_b=0$ and action $a=0$ is taken (i.e., no request is accepted for using slices), then the state of BE queue will transit to $s_b=1$ with probability $p_n^b$ and stay at state $s_b=0$ with probability $(1-p_n^b)$. 
	
	Similarly, we can define the matrix  ${\mathbf{B}}_{b,b'}(a)$ as follows:
	\begin{equation}
	{\mathbf{B}}_{b,b'}(a) =	\!  \left[\!	\begin{array}{c@{\hspace{0.3em}}c@{\hspace{0.3em}}c@{\hspace{0.3em}}c}
	{\mathbf{G}}_{0,0}(a)	&	{\mathbf{G}}_{0,1}(a) &	\ldots & {\mathbf{G}}_{0,Q_g}(a) 	\\
	{\mathbf{G}}_{1,0}(a)	&	{\mathbf{G}}_{1,1}(a)	&	\ldots  & {\mathbf{G}}_{1,Q_g}(a)	\\
	\vdots &	\vdots	&	\ddots	&	\vdots		\\
	{\mathbf{G}}_{Q_g,0}(a) & {\mathbf{G}}_{Q_g,1}(a)	&	\ldots	&	{\mathbf{G}}_{Q_g,Q_g}(a)
	\end{array}\!	\right]
	\begin{array}{l}	\! \leftarrow	g=0	\\	\! \leftarrow	g=1	\\	\vdots	\\	\! \leftarrow	g = Q_g \end{array}		
	\label{eq:pmatrix_Q}
	\end{equation}
	where each row of matrix ${\mathbf{B}}_{b,b'}(a)$ corresponds to the number of requests in the GS queue, and ${\mathbf{G}}_{g,g'}(a)$ is defined by:
	\begin{equation}
	{\mathbf{G}}_{g,g'}(a) =	\!  \left[\!	\begin{array}{c@{\hspace{0.3em}}c@{\hspace{0.3em}}c@{\hspace{0.3em}}c}
	{\mathbf{R}}_{0,0}(a)	&	{\mathbf{R}}_{0,1}(a) &	\ldots & {\mathbf{R}}_{0,R}(a) 	\\
	{\mathbf{R}}_{1,0}(a)	&	{\mathbf{R}}_{1,1}(a)	&	\ldots  & {\mathbf{R}}_{1,R}(a)	\\
	\vdots &	\vdots	&	\ddots	&	\vdots		\\
	{\mathbf{R}}_{R,0}(a) & {\mathbf{R}}_{R,1}(a)	&	\ldots	&	{\mathbf{R}}_{R,R}(a)
	\end{array}\!	\right]
	\begin{array}{l}	\! \leftarrow	r=0	\\	\! \leftarrow	r=1	\\	\vdots	\\	\! \leftarrow	r = R \end{array}		
	\label{eq:pmatrix_R}
	\end{equation}
	where each row of matrix ${\mathbf{G}}_{g,g'}(a)$ corresponds to the state of radio resources, and ${\mathbf{R}}_{r,r'}(a)$ can be defined as follows:
	\begin{equation}
	{\mathbf{R}}_{r,r'}(a) =	\!  \left[\!	\begin{array}{c@{\hspace{0.3em}}c@{\hspace{0.3em}}c@{\hspace{0.3em}}c}
	{\mathbf{C}}_{0,0}(a)	&	{\mathbf{C}}_{0,1}(a) &	\ldots & {\mathbf{C}}_{0,C}(a) 	\\
	{\mathbf{C}}_{1,0}(a)	&	{\mathbf{C}}_{1,1}(a)	&	\ldots  & {\mathbf{C}}_{1,C}(a)	\\
	\vdots &	\vdots	&	\ddots	&	\vdots		\\
	{\mathbf{C}}_{C,0}(a) & {\mathbf{C}}_{C,1}(a)	&	\ldots	&	{\mathbf{C}}_{C,C}(a)
	\end{array}\!	\right]
	\begin{array}{l}	\! \leftarrow	c=0	\\	\! \leftarrow	c=1	\\	\vdots	\\	\! \leftarrow	c = C \end{array}		
	\label{eq:pmatrix_C}
	\end{equation}
	where each row of matrix ${\mathbf{R}}_{r,r'}(a)$ corresponds to the state of computing resources, and ${\mathbf{C}}_{c,c'}(a)$ can be defined as follows:
	\begin{equation}
	{\mathbf{C}}_{c,c'}(a) =	\!  \left[\!	\begin{array}{c@{\hspace{0.3em}}c@{\hspace{0.3em}}c@{\hspace{0.3em}}c}
	{p}_{0,0}(a)	&	{p}_{0,1}(a) &	\ldots & {p}_{0,\Delta}(a) 	\\
	{p}_{1,0}(a)	&	{p}_{1,1}(a)	&	\ldots  & {p}_{1,\Delta}(a)	\\
	\vdots &	\vdots	&	\ddots	&	\vdots		\\
	{p}_{\Delta,0}(a) & {p}_{\Delta,1}(a)	&	\ldots	&	{p}_{\Delta,\Delta}(a)
	\end{array}\!	\right]
	\begin{array}{l}	\! \leftarrow	\delta=0	\\	\! \leftarrow	\delta=1	\\	\vdots	\\	\! \leftarrow	\delta = \Delta \end{array}		
	\label{eq:pmatrix_S}
	\end{equation}
	where each row of matrix ${\mathbf{C}}_{c,c'}(a)$ corresponds to the state of storage resources. Each element ${p}_{\delta,\delta'}(a)$ represents the state transition probability of the storage resource from state $\delta$ to $\delta'$ when action $a$ is taken at state $\delta$. 
	
	Here, we note that different from matrices ${\mathbf{P}}(a)$ and ${\mathbf{B}}_{b,b'}(a)$ where transition probabilities depend on the arrival process of requests and the action of the Network Orchestrator, the transition probabilities of matrices ${\mathbf{G}}_{g,g'}(a)$, ${\mathbf{R}}_{r,r'}(a)$, and ${\mathbf{C}}_{c,c'}(a)$, depend on the departure process of requests and the action of the Network Orchestrator.

	\begin{figure*}[!]
		\begin{center}
			\epsfxsize=6.2 in \epsffile{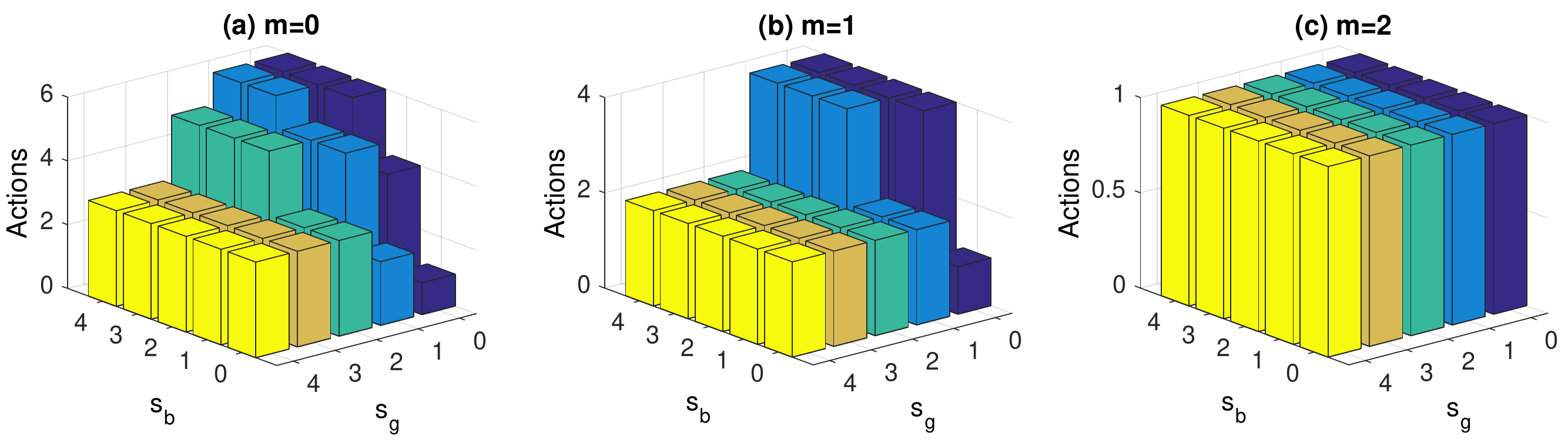}  
			\caption{Cross-slice Orchestrator Optimal Policy as a function of the GS ($s_g$) and BE ($s_b$) queues for (a) m=0, (b) m=1, and (c) m=2 deployed slices.}
			\label{fig:Plot_Optimal_Policy}
		\end{center}
	\end{figure*}
	
	\subsection{Reward Function}
	The proposed solution aims to maximize the revenue of the provider in term of admitted slice requests. In particular, if we denote $r_b$ and $r_g$ as the rewards (e.g., monetary values) which the provider receives from the BE and GS services clients if the provider serves a BE and GS request, respectively. Then, the immediate reward function can be defined as follows:
	\begin{equation}
	R(t) = a_g(t) r_g + a_b(t) r_b	.
	\end{equation}
	
	The goal is to choose an optimal policy $\pi^*$ that maximizes the expected discounted sum over an infinite horizons:
	\begin{equation}
	\max_{\pi^*} \mathcal{R} = 	\sum_{t=0}^{\infty} \gamma^t R(s_t,\pi^*(s_t)) ,
	\end{equation}
	where $\gamma$ is the discount factor that satisfies $\gamma \in (0,1]$.

	\subsection{Value Iteration Algorithm}
	To find the optimal cross-slice orchestration policy, we adopt the value iteration algorithm~\cite{Puterman1994MDP}. In particular, the value iteration algorithm is an iterative procedure which calculates the expected optimal value of each state. Value iterations stop when the values calculated on two successive steps are close enough, i.e., 
	\begin{equation}
	\max_{s} |\mathcal{V}_{k}(s) - \mathcal{V}_{k-1}(s)| < \epsilon, \forall s \in \mathcal{S}
	\end{equation}
	where $\epsilon$ is a predefined threshold value. The smaller $\epsilon$ is, the higher the precision of the algorithm is. The value iteration algorithm then can be expressed as in Algorithm~\ref{algorithm}:
	
	\begin{algorithm}
		\caption{Value iterative algorithm to obtain the optimal policy for the provider.}
		\label{algorithm}
		\textbf{1. Given:} \\
		\phantom{5}	1) Transition probability matrix ${\mathbf{P}}$ and reward function $R$. \\
		\phantom{5}	2) Initiate the state value vector $\mathbf{V}_{0}=\mathbf{0}$. \\	
		\textbf{2. Iteration:} 	\\
		\textbf{Repeat} \\
		\phantom{5}		\textbf{For} each state $s$, do for each action $a$ \\
		\phantom{10}			$Q_{k}(s,a)$ = $R(s,a) + \gamma \sum_{s'} \mathbf{P}_{s,s'}(a) \mathbf{V}_{k-1}(s) $ \\
		\phantom{10}			$\pi_{k}^{*}(s)$ = $\argmax_{a} Q_{k}(s,a)$ \\
		\phantom{10}			$\mathcal{V}_{k}(s)$ = $Q_{k}(s,\pi_{k}^{*}(s))$ \\
		\phantom{5}		\textbf{end} \\
		\textbf{Until} $|\mathcal{V}_{k}(s) - \mathcal{V}_{k-1}(s)| < \epsilon, \forall s \in \mathcal{S}$	\\
		\textbf{3. Return:} 	\\
		$\pi^{*}=\big[\pi^{*}(1), \ldots, \pi^{*}(s), \ldots, \pi^{*}(|S|) \big]^{\top}$.
	\end{algorithm}
	
	In Algorithm~\ref{algorithm}, $\mathbf{V}_k=\big[\mathcal{V}_k(1), \ldots, \mathcal{V}_k(|S|) \big]^{\top}$, where $\mathcal{V}_k(s)$ is the value of state $s \in \mathcal{S}$ at loop-k, $|S|$ is the total number of states in the state space $\mathcal{S}$, and $^\top$ is the transpose function.

	\subsection{Performance Analysis}
	
	After obtaining the optimal policy $\pi^*$, we can derive the steady state probability of the system, i.e., $\phi$, by solving the following equation:
	\begin{equation}
	\phi {\mathbf{P}}(\pi^*) = \phi 	,
	\end{equation}
	where ${\mathbf{P}}(\pi^*)$ is the transition probability matrix of the system under the optimal policy $\pi^*$. Here, we note that $\phi = \big[\phi(1), \ldots, \phi(s), \ldots, \phi(|S|) \big]^{\top}$ and $\sum_{s \in \mathcal{S}} \phi(s) = 1$. Then, the average reward of the system and the dropping probabilities of requests can be calculated as follows.
	
	\begin{itemize}
		\item {\em Average reward:} The average reward of the provider can be obtained from
		\begin{equation}
		\mathcal{R} = \sum_{s \in {\mathcal{S}}}  \phi (s) R(s,\pi^*(s)).
		\label{utilization} 
		\end{equation}
		\item {\em Dropping probability:} Dropping probability is the probability in which a slice request arrives to the system and is discarded because the queue is full. The dropping probabilities of GS and BE requests, i.e., $P_{d}^{g}$ and $P_{d}^{b}$, respectively, can be obtained as follows:
		\begin{equation}
		P_{d}^{g} = p_n^g \sum_{s \in \mathcal{S}_1} \phi (s)	, \phantom{5}
		P_{d}^{b} = p_n^b \sum_{s \in \mathcal{S}_2} \phi (s)	,
		\end{equation}	
		where $\mathcal{S}_1$ and $\mathcal{S}_2$ are sets of states in which the GS and BE queues are full, respectively. 
	\end{itemize}

	\section{Performance Evaluation} 
	\label{sec:Per_Eva}

	\subsection{Parameter Setting}
	
	We consider a 5G network in which the maximum number of radio (R), computing (C), and storage ($\Delta$) resources are set at 4 units. Each request from both GS and BE queues will require d=2 units of radio, computing, and storage resources. The maximum number of arriving BE requests is  $N_b=1$, the maximum BE queue length is $Q_b= 4$, the arrival and departure probabilities of one BE request are set at $p_n^b=p_l^b= 0.85$, and the immediate reward to allocate one slice to one BE request is $r_b= 1$ unit. The maximum number of arriving GS requests is $N_g= 1$, the maximum GS queue length is $Q_g= 4$, the arrival and departure probabilities of one GS request are set equal to $p_n^g=p_l^g= 0.35$, and the immediate reward to allocate one slice to one BE request is $r_g= 1.553$ unit. In this way, we aim to model the fact that the GS requests are more sporadic than that of the BE ones, but they can potentially provide higher revenues to the provider. The arrival and departure probabilities of requests will be varied to evaluate the performance of the proposed solution under different circumstances. For the value iteration algorithm, the discount factor $\gamma$ is set at $0.9$.
	
	\begin{figure*}[!]
		\begin{center}
			$\begin{array}{ccc}
			\epsfxsize=2.0 in \epsffile{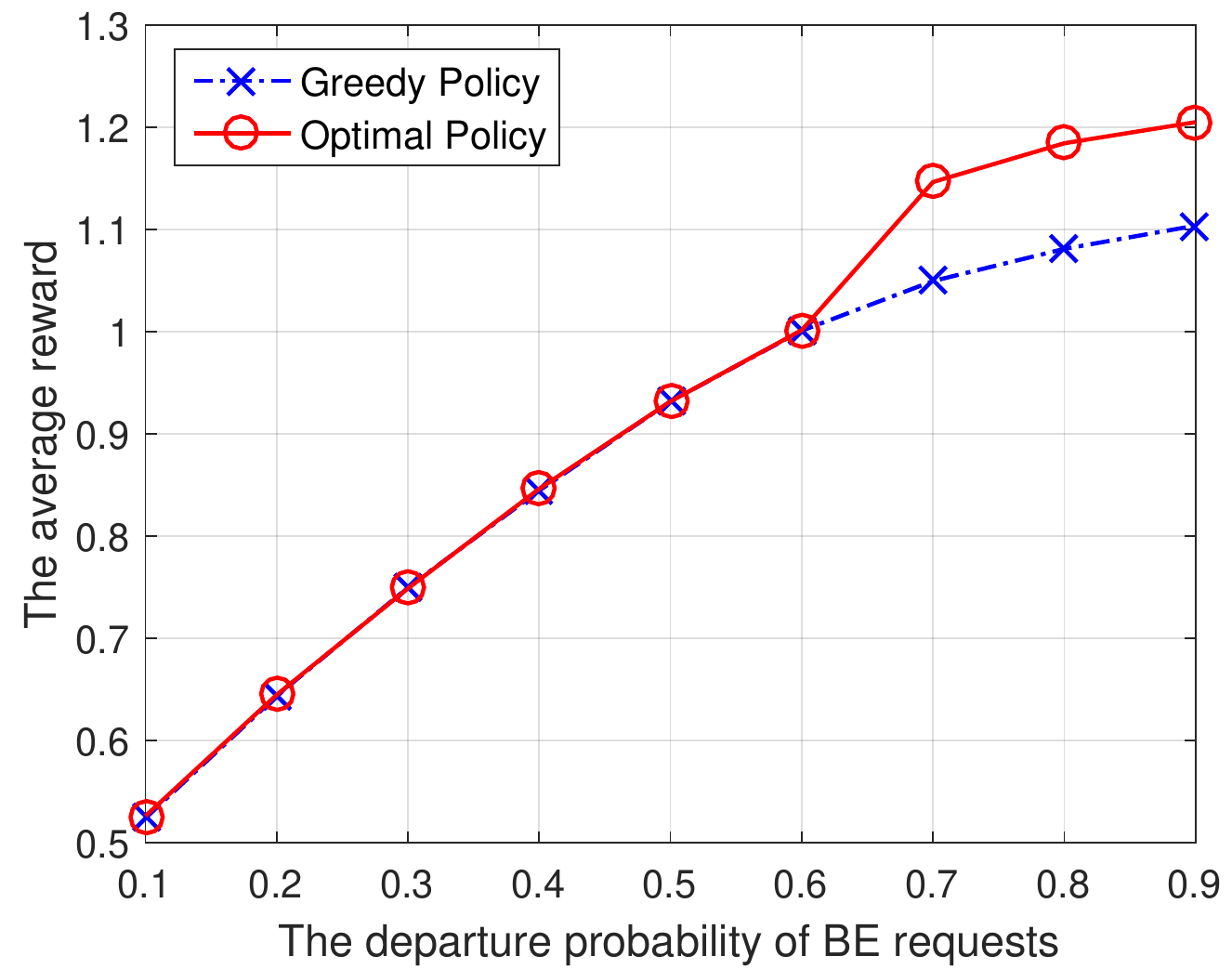}    &
			\epsfxsize=2.0 in \epsffile{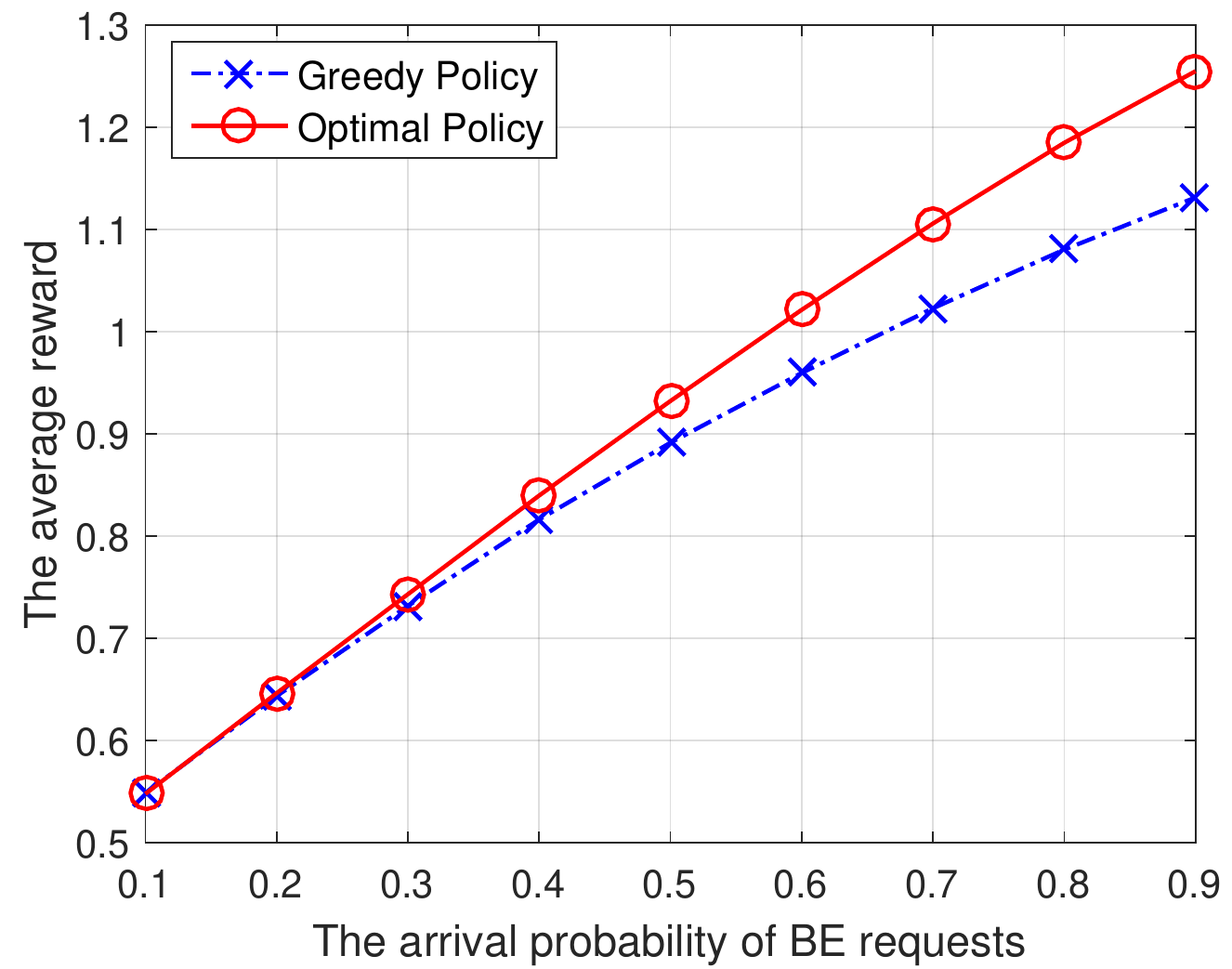}    &
			\epsfxsize=2.0 in \epsffile{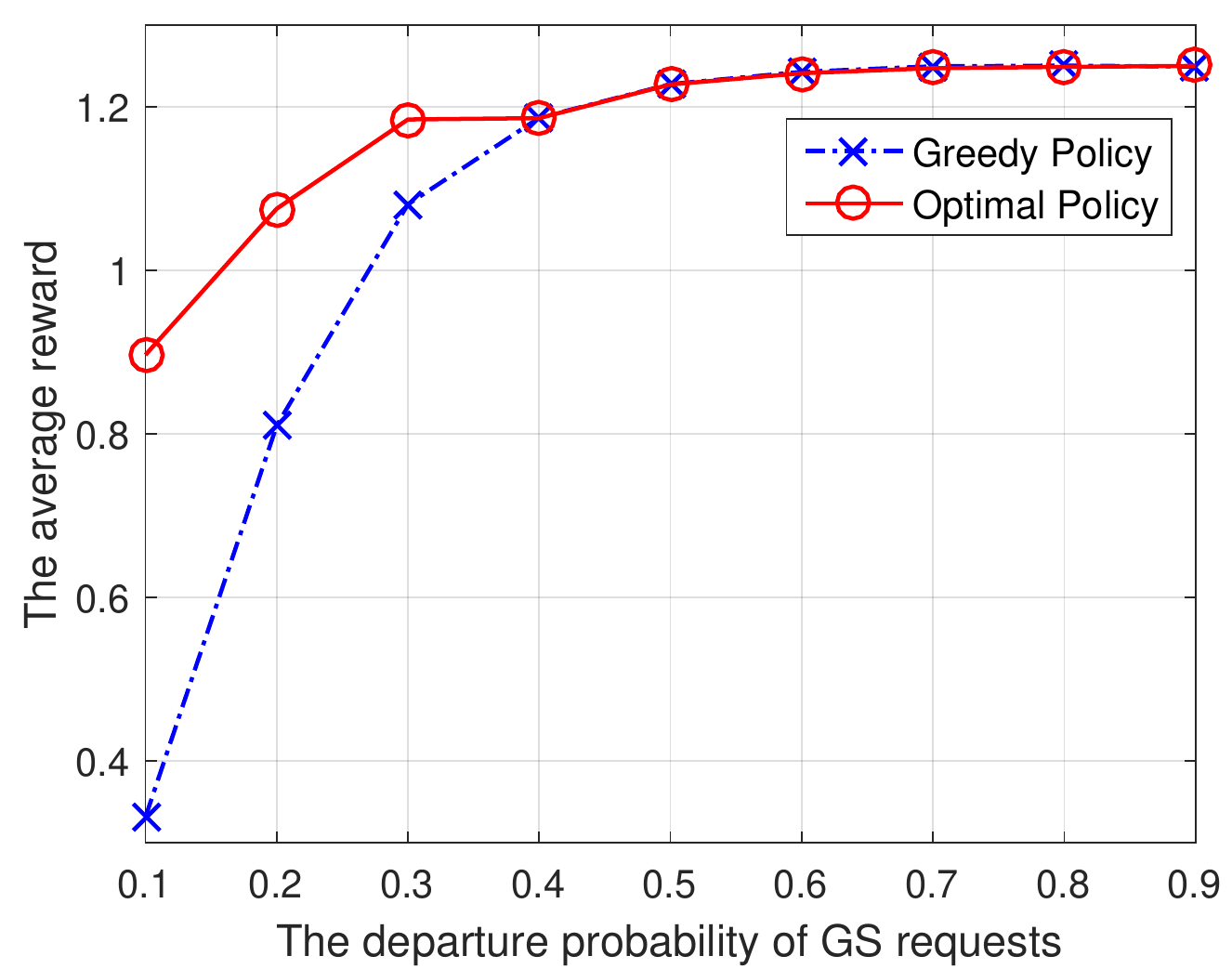} \\ [-0.2cm]
			(a)  & (b) & (c) 
			\end{array}$
			\caption{Average reward with the optimal and greedy policies as a function of the departure probability of BE requests, the arrival probability  of BE requests, and the departure probability of GS requests.}
			\label{fig:average_reward}
		\end{center}
	\end{figure*}

	\subsection{Numerical Results}
	
	\paragraph{Optimal Policy}

	In Fig.~\ref{fig:Plot_Optimal_Policy}, we show the optimal policy of the Network Orchestrator obtained by the Algorithm~\ref{algorithm}. Here, we denote $m$ as the current number of deployed slices. Given the parameter setting, the maximum number of slices is 2, and thus we have three cases corresponding to the cases when $m=0$, $m=1$, and $m=2$ (note that having the number of deployed slices corresponds to observe the resource availability). Since each slice request requires 2 units of radio, computing, and storage resources, and the maximum number of each of these resources is set to 4, we have $a_b, a_g \in \{0,1,2\}$. Thus, there are 6 actions in the action space in this case, i.e., $a=\{1,2,3,4,5,6\}$ corresponding to $(a_b,a_g)=\{(0,0),(0,1),(0,2),(1,0),(1,1),(2,0)\}$, respectively. 
	
	In Fig.~\ref{fig:Plot_Optimal_Policy} (a), when $m=0$ (i.e., there is no active slice in the system), if the number of requests in the GS queue is large, e.g., $s_g=3$ or $4$, the Cross-Slice Orchestrator will accept requests waiting in the GS queue as many as possible ($a=3$). However, when the number of requests in the GS queue is small, e.g., $s_g=1$ or $2$, and the number of requests in BE queue is large, the Cross-Slice Orchestrator will accept one request from BE queue and one request from the GS queue ($a=5$). When the number of requests in the GS queue is very small and the number of requests in the BE queues is very large, the Cross-Slice Orchestrator will accept requests in the BE queue as many as possible ($a=6$). In Fig.~\ref{fig:Plot_Optimal_Policy} (b), when $m=1$ (i.e., there is one request using one slice in the system), the Cross-Slice Orchestrator will accept requests from the BE queue ($a=4$) only when the GS queue is empty or when there is only one request in the GS queue and there are more than 2 requests in the BE queue. Otherwise, the Cross-Slice Orchestrator will choose a request from the GS queue ($a=2$). Finally, in Fig.~\ref{fig:Plot_Optimal_Policy} (c), when $m=2$ (i.e., there are 2 active requests using slices in the system), the Cross-Slice Orchestrator will accept no request ($a=1$). As expected, the obtained optimal policy implies that the requests with higher rewards have greater opportunities to be allocated network resources.

	\paragraph{Performance Evaluation}

	\begin{figure*}[!]
		\begin{center}
			$\begin{array}{ccc}
			\epsfxsize=2.0 in \epsffile{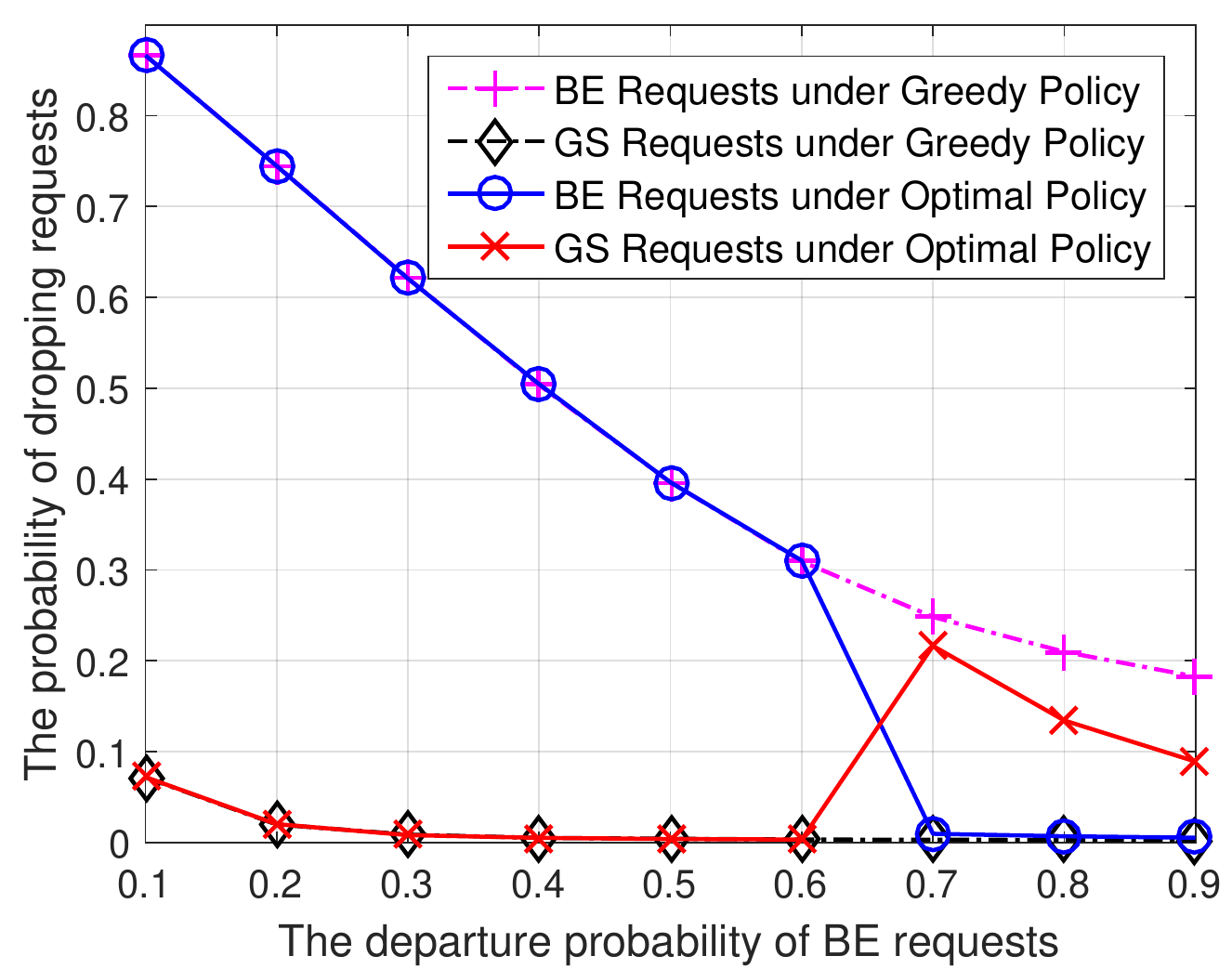}    &
			\epsfxsize=2.0 in \epsffile{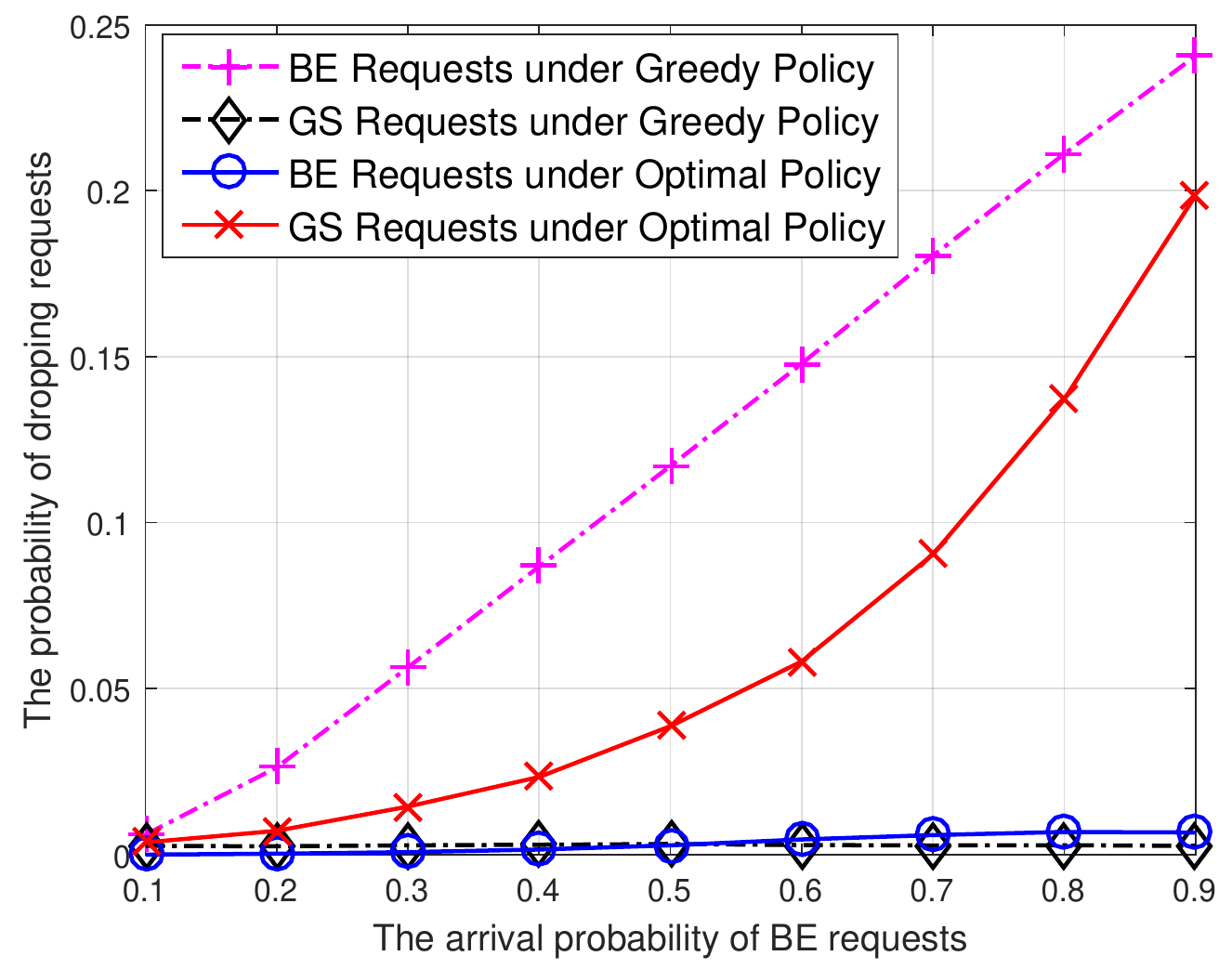}    &
			\epsfxsize=2.0 in \epsffile{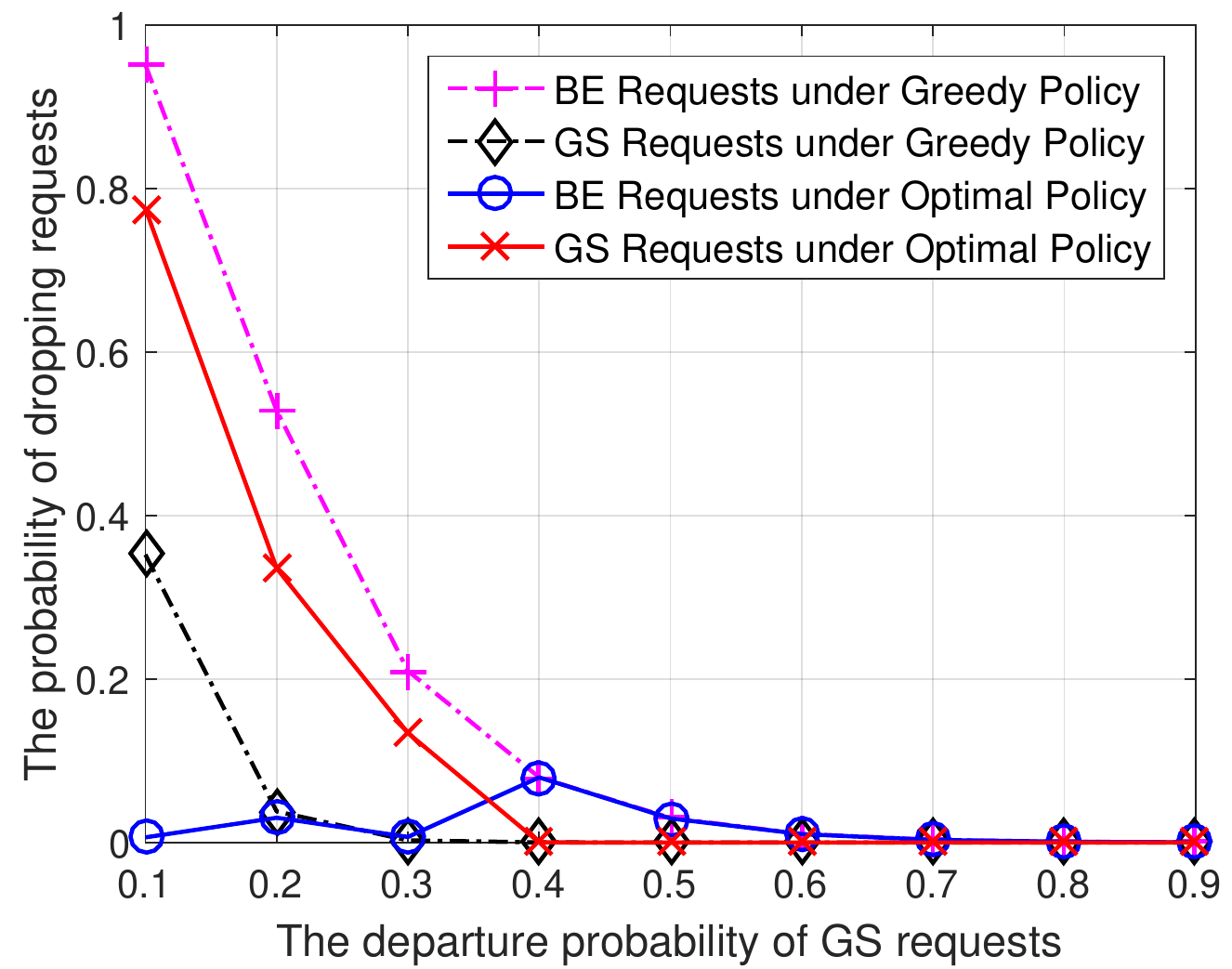} \\ [-0.2cm]
			(a)  & (b) & (c) 
			\end{array}$
			\caption{Request dropping probability with the optimal and greedy policies as a function of the departure probability of BE requests, the arrival probability  of BE requests, and the departure probability  of GS requests.}
			\label{fig:dropping_prob}
		\end{center}
	\end{figure*}

	We now vary the departure probability of GS requests, the departure probability of BE requests, and the arrival probability of BE requests to evaluate the performance of the Cross-Slice Orchestrator in terms of the average reward and the dropping probability of requests. In this case, we will compare the results obtained from the optimal solution with the results obtained from the greedy policy. For the greedy policy, the Cross-Slice Orchestrator chooses the action that maximizes its immediate reward. 
	
	In Fig.~\ref{fig:average_reward} (a) and Fig.~\ref{fig:dropping_prob} (a), as the departure probability of BE requests increases, the average reward will increase and the dropping probability will decrease for both policies. When the departure probability of the BE requests is low, e.g., lower than $0.6$, the GS requests will have higher priorities since they have higher rewards, and thus the optimal policy is the same as the greedy policy, i.e., accept as many GS requests as possible. However, when the departure probability of BE requests is high, the BE requests will be preferable since given a fixed time period, more BE requests can be served than GS requests, yielding a higher overall reward for the provider. As a result, the average reward obtained by the optimal policy will be higher than that of the greedy policy when the departure probability of BE requests is high.
	
	In Fig.~\ref{fig:average_reward} (b) and Fig.~\ref{fig:dropping_prob} (b), we vary the request arrival probability of BE requests and evaluate the average reward and the dropping probability of the optimal policy. When the arrival probability of BE requests is lower than $0.3$, the optimal policy is the greedy policy because now the system is able to serve all requests arriving at the system, and thus the average rewards obtained by both policies are the same. However, when the arrival probability of BE requests is higher than $0.3$, the system does not have sufficient resources to serve all incoming requests. Thus, the average reward obtained by the optimal policy is greater than that of the greedy policy since the optimal policy can balance between the immediate and the long-term rewards. 
	
	In Fig.~\ref{fig:average_reward} (c), when the departure probability of the GS request is $0.1$, the average reward obtained by the optimal policy is nearly $2.8$ times greater than that of the greedy policy. The reason is that when the departure probability is very low, if the GS requests are always accepted, there will be no opportunity for BE requests to be served, and thus the dropping probability of BE requests is very high, i.e., 0.78 (as shown in Fig.~\ref{fig:dropping_prob} (c)). However, for the optimal policy, the Cross-Slice Orchestrator will balance BE and GS requests to achieve the best performance. As a result, the average reward obtained by the optimal policy is always greater than that of the greedy policy when the departure probability of GS requests is low, i.e., lower than 0.4. When the departure probability of GS is high, the optimal policy will accept GS request as many as possible, and thus the performances of the optimal policy and greedy policy are identical. Results from Fig.~\ref{fig:average_reward} and Fig.~\ref{fig:dropping_prob} reveal that arrival and departure probabilities of requests are also important factors which impact the optimal decision and the performance of the system.
	

	\section{Summary} 
	\label{sec:Sum}
	
	In this paper, we have introduced a system model which allows the 5G network provider to provide slice-as-a-service in a dynamic fashion based on the service requirements and the resource availability. We have then formulated the cross-slice admission control and resource allocation optimization problem as the Markov decision process, and applied the value iteration algorithm to find the corresponding optimal policy. Simulation results have clearly shown that the proposed solution can help the provider to maximize its revenue given its resource constraints and the service requirements. In the future, we will study online learning methods with linear function approximation to deal with the curse-of-dimensionality and the curse-of-model problems in 5G networks.

	\bibliographystyle{IEEE}

\end{document}